\documentclass[aps,prl,twocolumn,groupedaddress,superscriptaddress]
{revtex4}
\usepackage{graphicx} 
\usepackage{dcolumn} 
\usepackage{longtable} 
\usepackage{amssymb} 
\usepackage{amsmath} 
\usepackage{amsfonts} 
\usepackage{slashed} 
\usepackage[dvipsnames]{xcolor} 
\usepackage{soul}
\usepackage{mathtools}
\usepackage[normalem]{ulem}

\newcommand{\be}{\begin{equation}} 
\newcommand{\ee}{\end{equation}}

\newcommand{\simle}{\hspace*{0.2em}\raisebox{0.5ex}{$<$}
	\hspace{-0.8em}\raisebox{-0.3em}{$\sim$}\hspace*{0.2em}}

\begin{document}

\title{A renormalizable theory for not-so-light nuclei}

\author{L. Contessi}
\email{lorenzo@contessi.net}
\affiliation{Universit\'e Paris-Saclay, CNRS-IN2P3, IJCLab, 91405 Orsay, France}

\author{M. Sch\"{a}fer}
\email{m.schafer@ujf.cas.cz}
\affiliation{Nuclear Physics Institute of the Czech Academy of Sciences, \v{R}e\v{z} 25068, Czech Republic}

\author{A. Gnech}
\affiliation{Department of Physics, Old Dominion University, Norfolk, Virginia 23529, USA}
\affiliation{Theory Center, Jefferson Lab, Newport News, Virginia 23610, USA}

\author{A. Lovato}
\affiliation{Physics Division, Argonne National Laboratory, Argonne, IL 60439, USA}
\affiliation{Computational Science Division, Argonne National Laboratory, Argonne, IL 60439, USA}
\affiliation{INFN-TIFPA Trento Institute of Fundamental Physics and Applications, 38123 Trento, Italy}

\author{U. van Kolck}
\affiliation{European Centre for Theoretical Studies in Nuclear Physics and Related Areas (ECT*), Fondazione Bruno Kessler, 38123 Villazzano (TN), Italy}
\affiliation{Universit\'e Paris-Saclay, CNRS-IN2P3, IJCLab, 91405 Orsay, France}
\affiliation{Department of Physics, University of Arizona, Tucson, Arizona 85721, USA}

\begin{abstract} 
We present an improved 
action for Pionless Effective Field Theory (EFT). Previous formulations of renormalizable nuclear 
EFTs have encountered 
instabilities in systems with more than four nucleons. We resolve this issue
by introducing a finite interaction range at leading order, which is compensated for in perturbation theory at next-to-leading order. 
Calculated ground-state energies of $^4$He, $^6$Li, $^{12}$C, and $^{16}$O converge and agree with experiment within theoretical uncertainties.
This first successful implementation of systematic renormalizability 
beyond the lightest nuclei enables
not only applications to larger nuclei but also extensions to other 
EFTs in the strong-coupling regime.  
\end{abstract}

\maketitle

Properly renormalized effective field theories (EFTs) \cite{Hammer:2019poc} 
have successfully described nuclear systems with $A \le 5$ nucleons \cite{Bagnarol:2023crb} with systematic and efficient control over theoretical uncertainties and short-range dynamics.
However, they have encountered difficulties in producing stable ground states at leading order (LO) in heavier systems, in contrast with phenomenological approaches that eschew a consistent hierarchy of the effects of interactions on observables (``power counting'') in favor of an adjustable short-distance regulator.

For bosonic systems such as atomic $^4$He clusters, agreement 
between EFT and phenomenological potentials is already good at LO~\cite{Bedaque:1998km,Platter:2004he,Bazak:2016wxm} and improves order by order~\cite{Ji:2012nj,Bazak:2018qnu}. Saturation occurs~\cite{Carlson:2017txq} and yields reasonable results for the corresponding quantum liquid~\cite{deleon2022equationstatestronglyinteracting}. 
Yet, shell effects in fermionic systems have posed a challenge, as experimentally stable nuclei such as $^6$Li, $^{16}$O, and $^{40}$Ca have not been obtained at LO in either Pionless EFT~\cite{Stetcu:2006ey,Contessi:2017rww,Bansal:2017pwn} or Chiral EFT~\cite{Yang:2020pgi}. 
It is not inherently problematic for LO calculations to fail in predicting the weak stability of medium-light nuclei because their relatively large theoretical uncertainties leave room for higher-order corrections. Nonetheless, 
unstable poles 
in the continuum spectrum
are unsuitable as a basis for the perturbation theory required for renormalizability at subleading orders~\cite{Yang:2023wci}. As a result, the convergence behavior of these theories for $A \ge 6$ remains largely unexplored.

To ensure LO stability in multicomponent fermion systems, such as nuclei,
two scenarios may be considered: (a) a change in power counting, or (b) 
variation of LO within its uncertainty band. The first option has been explored in Chiral EFT via combinatorial enhancement of few-nucleon forces~\cite{Yang:2021vxa}, with mixed results. The second scenario can be implemented in two complementary ways. One is to fit the LO interaction directly to reproduce the stability of larger systems --- a valid strategy, as differences in fitted low-energy observables diminish at higher orders. 
However, this approach becomes increasingly demanding 
and obscures a possible failure of the theory as the number of particles and density increase. Alternatively, one can include at LO a subset of subleading interactions that preserves regulator independence and is later compensated for perturbatively. These ``improved actions'' have shown clear benefits in two-body systems~\cite{Contessi:2024vae} and small $^4$He clusters~\cite{Contessi:2023yoz}.

We adopt here a similar strategy for nuclei in Pionless EFT up to next-to-leading order (NLO). Pionless EFT describes nuclei in terms of nucleons only, with few parameters at low orders, most notably a three-body force at LO whose sole parameter sets the energy and distance scales for $A \ge 3$ bound states. We compute the ground-state energies of selected nuclei ($^4$He, $^6$Li, $^{12}$C, and $^{16}$O) and investigate their stability under various improvements. We also compare our results with experimental data to evaluate the potential of the approach for practical many-body calculations. The good agreement we find suggests that 
a significant number of nuclear ground states lie within the regime of Pionless EFT, whose breakdown scale is comparable to the pion mass~\cite{Ekstrom:2024dqr}. We thus provide, for the first time, a renormalized nuclear EFT that explains nuclear binding through a minimal set of essential, nonperturbative interactions, with details treated as perturbations.

Pionless EFT consists of contact interactions which, upon regularization with a momentum cutoff $\Lambda$, are represented by products of smeared Dirac delta functions $\delta_\Lambda(\vec{r}_{ij})$, where $r_{ij}$ is the distance between particles $i$ and $j$. 
We employ 
\begin{equation} 
\delta_\Lambda(\vec{r}) = \left(\frac{\Lambda}{2 \sqrt{\pi}}\right)^3~\exp\left(-\frac{\Lambda^2}{4} \vec{r}^{,2}\right), 
\label{eq:regulator} 
\end{equation}
although observables remain independent of the specific regulator form. Each interaction term contains a low-energy constant (LEC), whose dependence on $\Lambda$ is adjusted such that the residual cutoff dependence of observables can be made arbitrarily small as $\Lambda$ increases. In what follows, the cutoff dependence of the LECs is left implicit. Each LEC is fixed by fitting to a single experimental input. We perform the fits by solving the Schr\"odinger equation using the Numerov method 
in distorted-wave 
perturbation theory for two-body scattering, and the Stochastic Variational Method (SVM)~\cite{Suzuki:1998bn} for bound-state observables.

We follow the standard power counting for isospin-symmetric interactions, which has been thoroughly tested in few-body systems~\cite{Hammer:2019poc}. Electromagnetic isospin breaking terms are included, while we neglect contributions from the down–up quark-mass difference, as they enter at NLO and N$^2$LO, respectively, except in very low-energy scattering~\cite{Konig:2015aka}. 

LO includes two- and three-body potentials,
\begin{align}
    V^{(0)}_{\text{2B}}
    &=  \sum_{i<j} \left(C_{0,s}^{(0)} \mathcal{P}_{ij,s} 
    +C_{0,t}^{(0)} 
    \mathcal{P}_{ij,t} \right)  \delta_\Lambda(\vec{r}_{ij}),
    \label{eq:EFT_LO_2b}
    \\
    V^{(0)}_{\text{3B}}&= D^{(0)} \sum_{i<j<k} \sum_{\text{cyc}} \delta_\Lambda(\vec{r}_{ij})\,\delta_\Lambda(\vec{r}_{jk}),
    \label{eq:EFT_LO_3b}
\end{align}
where $\mathcal{P}_{ij,s}$ and $\mathcal{P}_{ij,t}$ are spin-isospin projectors in the $s$ (spin $0$, isospin $1$) and $t$ (spin $1$, isospin $0$) two-nucleon channels. The LECs $C_{0,s}^{(0)}$ and $C_{0,t}^{(0)}$ capture the most important low-energy $A=2$ physics~\cite{Chen:1999tn} encoded in the corresponding $S$-wave scattering lengths, $a_s=-18.95$ fm and $a_t=5.4112$ fm~\cite{Hackenburg:2006qd}. 
The LEC $D^{(0)}$ is needed for renormalization with $A\ge 3$~\cite{Bedaque:1999ve} and is calibrated to the triton binding energy, $E(^3\text{H})=-8.482$ MeV~\cite{Wang_2012}. 

At NLO, the theory is refined through the addition in first-order perturbation theory of two- and three-body corrections, as well as a four-body force:
\begin{eqnarray}
    V_{\text{2B}}^{(1)}
    &=& \sum_{i<j} \left(
    C_{0,s}^{(1)}+C_{1,s}^{(1)}r_{ij}^{2}\right)  \mathcal{P}_{ij,s} \, \delta_\Lambda(\vec{r}_{ij})
    \nonumber \\
    &+& \sum_{i<j} \left(
    C_{0,t}^{(1)}+C_{1,t}^{(1)}r_{ij}^{2}\right)  \mathcal{P}_{ij,t} \, \delta_\Lambda(\vec{r}_{ij})   
    \nonumber \\
    &+& \sum_{i<j} \left( C_{pp}^{(1)} \mathcal{P}_{p_ip_j} \delta_\Lambda(\vec{r}_{ij}) 
    +\frac{e_ie_j}{4\pi r_{ij}}\right),
    \label{eq:NLO_2b} \\
    V_{\text{3B}}^{(1)}
    &=&D^{(1)} \sum_{i<j<k} \sum_{\text{cyc}} \delta_\Lambda(\vec{r}_{ij})\,\delta_\Lambda(\vec{r}_{jk}),
    \label{eq:NLOa_3b}\\
    V_{\text{4B}}^{(1)}
    &=& E^{(1)} \sum_{i<j<k<l}\, \sum_{\text{cyc}}  \delta_\Lambda(\vec{r}_{ij})\, \delta_\Lambda(\vec{r}_{jk}) \, \delta_\Lambda(\vec{r}_{kl}).
    \label{eq:NLOa_4b}
\end{eqnarray} 
The two-body LECs $C_{0,s}^{(1)}$, $C_{1,s}^{(1)}$, $C_{0,t}^{(1)}$, and $C_{1,t}^{(1)}$ are adjusted to reproduce the two-body effective ranges $r_s=2.750$ fm and $r_t=1.753$ fm \cite{Hackenburg:2006qd} while keeping the scattering lengths fixed. 
The local form of NLO interactions is equivalent to the two-derivative form that results directly from the EFT Lagrangian \cite{Contessi:2023yoz}. Renormalization of the Coulomb interaction between protons of charge $e_i=e$ ($e_i=0$ for neutrons) requires \cite{Konig:2015aka} an additional LEC $C_{pp}^{(1)}$ in the proton-proton $s$ channel, onto which one projects with $\mathcal{P}_{p_ip_j}$. The NLO two-body corrections shift the triton energy, which is restored to the correct value by adjusting $D^{(1)}$. Additionally, they split helion and triton. When $C_{pp}^{(1)}$ is determined from two-nucleon scattering, the helion-triton splitting is reproduced within the NLO uncertainty \cite{Konig:2015aka}. For simplicity, we instead calibrate $C_{pp}^{(1)}$ with the experimental helion energy $E(^3\text{He})=-7.718$ MeV \cite{Wang_2012}. The four-body LEC $E^{(1)}$, required for renormalization \cite{Bazak:2018qnu}, is calibrated to the 
alpha-particle binding energy, $E(^4\text{He})=-28.3$ MeV \cite{Wang_2012}.

The theory constructed above is not practical for calculations involving nuclei larger than $^4$He due to instability issues. Most ground states lie close to the breakup threshold into clusters, as the observed binding energy per nucleon, $-E_A/A$, typically differs by less than 15\% from that of $^4$He. In contrast, the expansion parameter $\xi_A$ in Pionless EFT is expected to be larger, though it remains highly uncertain. A crude estimate is given by the ratio between the characteristic momentum of the system and the breakdown scale of the theory, $\xi_A \sim Q_A/M_{\rm hi}$. If each nucleon of mass $m$ contributes similarly to the total energy, then $Q_A \sim \sqrt{-2 m E_A/A}$. For atomic $^4$He systems, using the inverse van der Waals length as the breakdown scale leads to a significant overestimate of the expansion parameter~\cite{Bazak:2016wxm, Bazak:2018qnu, Contessi:2023yoz}. In nuclear systems, the breakdown scale may exceed the naive estimate of the pion mass $m_\pi$, potentially reaching $M_{\rm hi} \sim 1.4\, m_\pi$~\cite{Ekstrom:2024dqr}. Even with this higher estimate, $\xi_A$ can still exceed 50\%, making the LO uncertainty larger than the relative separation between the ground-state energy and the breakup threshold.

To obtain stability at LO, we improve the theory by introducing artificial 
ranges into the LO interactions, constrained to lie within the LO uncertainty and be removable through higher-order perturbative corrections. This strategy was demonstrated for atomic $^4$He clusters in Ref.~\cite{Contessi:2023yoz}. In the nuclear case, we introduce three separate fake ranges: one for each two-body interaction ($xR_s$, $xR_t$) and one for the three-body interaction ($xR_3$). The improved LO potentials are then
\begin{eqnarray}
    \tilde{V}^{(0)}_{\text{2B}}(x)
    &=&  \tilde{C}_{0,s}^{(0)} \sum_{i<j} \mathcal{P}_{ij,s}  \delta_{(x R_s)^{-1}}(\vec{r}_{ij})    
    \nonumber\\
    &+&  \tilde{C}_{0,t}^{(0)}  \sum_{i<j} \mathcal{P}_{ij,t} \delta_{(x R_t)^{-1}}(\vec{r}_{ij}),     
    \label{eq:EFT_LO_I_2B}\\
    \tilde{V}^{(0)}_{\text{3B}}(x)&=& \tilde{D}^{(0)} \!\! \sum_{i<j<k} \sum_{\text{cyc}} \delta_{(x R_3)^{-1}}(\vec{r}_{ij})\,\delta_{(x R_3)^{-1}}(\vec{r}_{jk}).
    \label{eq:EFT_LO_I_3B}
\end{eqnarray}
While the fake ranges could, in principle, be chosen independently, for simplicity we tie them together through a single parameter $x$, using fixed values for $R_s$, $R_t$, and $R_3$. We set $R_s = 0.8970$ fm and $R_t = 0.7719$ fm to match the two-body effective ranges; the parameter $x$ then measures the fraction of these ranges included at LO, generalizing to two channels the strategy of Ref.~\cite{Contessi:2024vae}. We choose $R_3 = 0.4149$ fm to reproduce the optimal $^4$He binding energy, $E^{(0)}(^{4}\text{He}) \simeq -29.57$ MeV at $x = 1$. This represents an
improvement over the $x = 0$ result, $E^{(0)}(^{4}\text{He}) \simeq -(31.77 \pm 0.01 \pm 6.6)$ MeV, extrapolated as described below 
with coefficient $q_4^{(0)} = 29.03 \pm 0.11$. Once the improvement is introduced, all LECs become $x$-dependent, but by construction, the LO (and only LO) predictions of the theory 
become cutoff independent. 

The choice of $x$ is constrained by the requirement that the LO 
improvement can be compensated by subleading orders, which we expect to be feasible for $x\lesssim 1$~\cite{Contessi:2023yoz, Contessi:2024vae}.
Two- and three-body improvements 
are directly related to 
contributions to
the effective ranges at NLO
and to the range of the three-body force
at N$^2$LO~\cite{Ji:2012nj}, respectively. 
Two-body effects 
are not expected to be larger
as long as the corresponding fake ranges 
are $\lesssim M_{\rm hi}^{-1}$, which is the case for $x \lesssim 1$. 
For three bodies, we have verified that the doublet neutron-deuteron scattering length $a_{1/2}(nd)$ remains in the narrow range 
0.70–0.72~fm when $R_3/\mathrm{fm}$ is varied between 0.50 and 0.33
with fixed $R_s$ and $R_t$ at $x=1$. The proximity to the experimental value $a_{1/2}(nd) = (0.65 \pm 0.04)$~fm~\cite{Dilg:1971dkn} indicates no issues arise unless the fake three-body range becomes significantly larger than the one we employ, which is roughly half the size of the two-body ranges.

We apply the improved action to $^4$He and $^6$Li using the SVM, and to $^{12}$C and $^{16}$O using the variational Monte Carlo method based on neural-network quantum states (VMC-NQS)~\cite{Adams:2020aax,Gnech:2021wfn,Gnech:2023prs}. VMC-NQS results have been benchmarked against SVM calculations in $A \leq 6$ nuclei. The VMC-NQS energies and their statistical errors are estimated as in Ref.~\cite{Gnech:2023prs}. In all these calculations, we treat the LO contributions nonperturbatively and the NLO corrections in first-order perturbation theory.

\begin{figure}
    \centering
    \includegraphics[width=\linewidth]{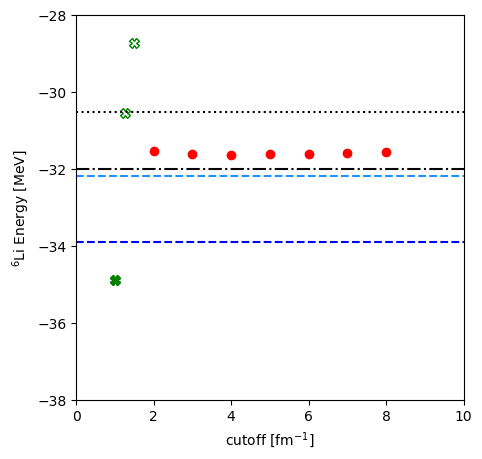}
    \caption{Ground-state energy of $^6$Li as a function of the cutoff. Green crosses represent unimproved LO (empty for unbound results), while the dark and light blue dashed lines indicate improved LO at $x=1$ and $x=0.9$, respectively. Red circles represent improved NLO results with $x=1$. The experimental threshold for $^4$He-$^2$H breakup is indicated by the black dotted line, while the experimental ground-state energy is given by the black dash-dotted line. Numerical errors are negligible compared to the thickness of the points and lines.}
    \label{fig:Li}
\end{figure}

\begin{figure}
    \centering
    \includegraphics[width=\linewidth]{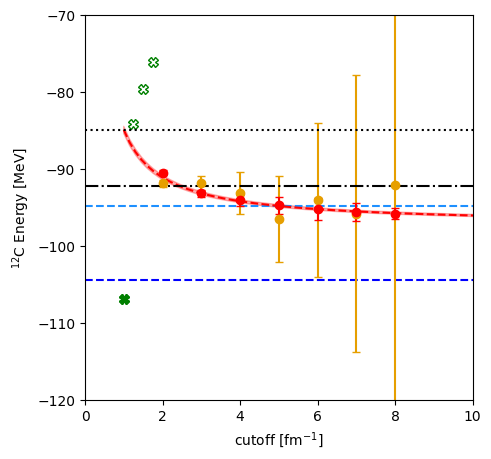}
    \caption{Ground-state energy of $^{12}$C 
    as a function of the cutoff. The red band is a fit to $x=1$ improved NLO points using Eq. \eqref{eq:cutoffextrapol}. Orange circles represent improved NLO results with $x=0.9$.  Other symbols as in Fig. \ref{fig:Li}, except that the black dotted line is the threshold for three-$^4$He breakup. Error bars represent statistical errors only.
    }
    \label{fig:C}
\end{figure}

\begin{figure}
    \centering
    \includegraphics[width=\linewidth]{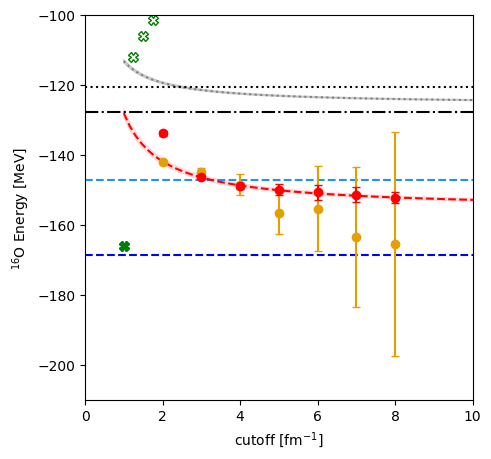}
    \caption{Ground-state energy of $^{16}$O 
    as a function of the cutoff. 
    Symbols as in Fig.~\ref{fig:Li}, except that the gray dash-dotted line represents the experimental $^4$He-$^{12}$C threshold, while the NLO theoretical threshold is indicated by a gray band around a dotted line.}
    \label{fig:O}
\end{figure}

Figures~\ref{fig:Li}, \ref{fig:C}, and \ref{fig:O} show the calculated ground-state energies of $^6$Li, $^{12}$C, and $^{16}$O as functions of $\Lambda$ in Eq.~\eqref{eq:regulator} for unimproved and improved LO, as well as for improved NLO. We explore values of $\Lambda$ up to 8~fm$^{-1}$, well beyond the expected EFT breakdown scale. The figures also display the experimental ground-state energies and the lowest experimental and NLO thresholds. Because the $^4$He energy is reproduced exactly at NLO, the experimental and NLO three-$^4$He thresholds for $^{12}$C coincide. Since the $^{12}$C energy is a prediction at NLO, the $^{12}$C–$^4$He threshold at NLO does not exactly match the experimental threshold. 
Although the $^2$H energy is not reproduced exactly,
the experimental and NLO thresholds for $^6$Li are nearly identical because the two-body shape parameters are small. The ground states computed at unimproved LO  --- Eqs.~\eqref{eq:EFT_LO_2b} and \eqref{eq:EFT_LO_3b} --- dissolve into the continuum at $\Lambda \lesssim 1$~fm$^{-1}$. This value is slightly smaller than the one observed for $^{16}$O in Ref.~\cite{Contessi:2017rww}, where the calibration of $C_{0,t}^{(0)}$ to the deuteron binding energy resulted in a stronger two-body interaction.

Our improved LO is cutoff-independent by construction and
numerical errors are smaller than the thickness of the lines. Renormalizability at NLO is demonstrated by extrapolating to large cutoffs using
\begin{equation}
E_A^{(1)}(\Lambda) = E_A^{(1)} \left(1+ \frac{q_A^{(1)}}{\Lambda}\right),
\label{eq:cutoffextrapol}
\end{equation}
where the parameter $q_A^{(1)}$ is set by physical scales and depends on $x$. We take the extrapolated $E_A^{(1)}$ as our central value. We quote two sources of uncertainty: 
numerical and truncation errors. 
The latter is estimated from the largest variation in energy observed above our lowest 
cutoff value
beyond the breakdown scale of the theory, $\Lambda = 2~\text{fm}^{-1}$. While this 
might be an underestimate, 
it provides a more quantitative assessment than naive dimensional analysis.

The improved LO results for $^6$Li in Fig.~\ref{fig:Li} yield a stable ground state with a binding energy about 10\% larger than experiment for $x = 1$. At NLO, the cutoff dependence is minimal ($q_6^{(1)} \simeq 0$) and the correction to the improved LO result is 
$\simeq 2$~MeV toward the experimental value. Taking the result at the highest cutoff as the central value, we obtain $E^{(1)}(^6\text{Li}) = -(31.57 \pm 0.02 \pm 0.3)$ MeV,
which compares well with the experimental value $E(^6\text{Li}) = -31.994$~MeV~\cite{Wang_2012} and supports the validity of the improved theory.

As shown in Fig.~\ref{fig:C}, the LO improvement at $x=1$ is even more manifest in $^{12}$C, where stability is a $\sim 25\%$ effect. The improved NLO results display cutoff stability well beyond the statistical uncertainties. Excluding the lowest cutoff value --- which may still be affected by higher-order $\Lambda^{-1}$ corrections --- we fit the NLO points using Eq.~\eqref{eq:cutoffextrapol}, yielding
$E^{(1)}(^{12}\text{C}) = -(97.3 \pm 0.1 \pm 5)$ MeV, with $q_{12}^{(1)} \simeq 25.0$~MeV. This result agrees with the experimental value $E(^{12}\text{C}) = -92.162$~MeV~\cite{Wang_2012}.

The instability problem is also resolved for $^{16}$O, where the improved LO and NLO results with $x=1$, shown in Fig.~\ref{fig:O}, lie well below the breakup thresholds, both experimental and theoretical at NLO. Extrapolating the NLO results yields $E^{(1)}(^{16}\text{O}) = -(155.6 \pm 0.3 \pm 20)$ MeV with $q_{16}^{(1)} \simeq 34.88$~MeV, in agreement with the experimental energy $E(^{16}\text{O}) = -127.619$~MeV~\cite{Wang_2012}.

We observe that stability at LO can be achieved for all three nuclei also for $x < 1$. However, reducing $x$ introduces significant numerical uncertainties at NLO. As shown in Figs.~\ref{fig:C} and \ref{fig:O}, the results for $x = 0.9$ agree with those for $x = 1$ within the (large) error bars, suggesting that stability is maintained in the relatively narrow range $0.9 \simle x \simle 1.0$. For $^6$Li, 
the system remains only weakly bound. The resulting expectation values are numerically unstable and cannot be reliably used in first-order perturbation theory. In the atomic case~\cite{Contessi:2023yoz}, it is possible to verify that the improvement remains perturbative with respect to the unimproved leading order over a broader range of $x$. Although we are more restricted here by the LO instability problem, there is no evidence suggesting that the same results could not be obtained through a careful application of the unimproved theory without altering the power counting. Still, within the limited class of improvements we investigated, $x = 1$ remains the optimal choice, as it minimizes numerical uncertainties while avoiding the risk of over-resummation. 

In all cases we examined, our improved NLO results lie closer to experiment than those at LO. The NLO energies exhibit stable cutoff behavior, consistent with a residual dependence $\propto \Lambda^{-1}$, indicating the robustness of the improved theory. The extrapolated energies for $^6$Li, $^{12}$C, and $^{16}$O agree with experiment within the estimated truncation error. The coefficient of the residual cutoff dependence is expected to scale as $q_A^{(1)} \sim \xi_A^2 M_{\rm hi}$; assuming minimal $x$ dependence, this gives $\xi_{12} \sim \xi_{16} \sim 0.4$. This value is smaller than the systematic uncertainty estimated via naive dimensional analysis, but not as small as the estimate inferred from $q_4^{(0)} \sim \xi_A M_{\rm hi}$. A similar situation is encountered in $^4$He atomic clusters~\cite{Bazak:2016wxm,Bazak:2018qnu,Contessi:2023yoz}, suggesting either an underestimation of the EFT breakdown scale or, more likely, an overestimation of the characteristic momentum in bound states. A more accurate estimate of the truncation error requires extensive N$^2$LO calculations, which are currently unavailable for $A>3$. Nevertheless, the improved NLO results support the applicability of Pionless EFT to heavier nuclei and provide strong motivation for advancing to higher orders.

Our specific improvement --- introducing fake ranges into the LO interactions --- brings our LO to a form that is essentially equivalent to the successful model of Refs.~\cite{Kievsky:2015dtk,Kievsky:2018xsl,Gattobigio:2019omi,Deltuva:2020aws,Gattobigio:2023fmo}. By providing stable ground states at LO as a foundation for perturbative corrections, the EFT framework enables higher-orders systematic improvements also beyond $A \le 4$ systems, distinguishing it from empirically successful models \cite{Kirscher:2010dgl,Bansal:2017pwn,Schiavilla:2021dun} that forgo renormalizability at the cost 
of a more limited expansion~\cite{Bub:2024gyz}.

The methodology developed in this work can 
be applied to improve the convergence rate and simplify numerical calculations in other 
EFTs,
such 
Chiral 
EFT~\cite{Yang:2020pgi}. Hence, this approach opens the door to model-independent {\it ab initio} calculations for a broader set of nuclei, potentially enabling large portions of the nuclear chart to be explored within a renormalizable theoretical framework.

\vspace{3mm}
\noindent
{\it Acknowledgments---}
The work of M.S. was supported by the Czech Science Foundation GA\v{C}R grant 22-14497S. A.G. acknowledges the support of Jefferson Lab that is sponsored by the U.S. Department of Energy under contract DE-AC05-06OR23177. A.L. is supported by the U.S. Department of Energy, Office of Science, Office of Nuclear Physics, under contract DE-AC02-06CH11357, by the 2020 DOE Early Career Award program, and by the NUCLEI SciDAC program. Numerical calculations were performed using an ISCRA award for accessing the LEONARDO supercomputer, owned by the EuroHPC Joint Undertaking, hosted by CINECA (Italy). We also acknowledge using resources of the National Energy Research Scientific Computing Center (NERSC), a Department of Energy Office of Science User Facility, 
through NERSC award NP-ERCAP0027079.
\bibliographystyle{apsrev}
\bibliography{NIA.bib}

\begin{thebibliography}{36}
\expandafter\ifx\csname natexlab\endcsname\relax\def\natexlab#1{#1}\fi
\expandafter\ifx\csname bibnamefont\endcsname\relax
  \def\bibnamefont#1{#1}\fi
\expandafter\ifx\csname bibfnamefont\endcsname\relax
  \def\bibfnamefont#1{#1}\fi
\expandafter\ifx\csname citenamefont\endcsname\relax
  \def\citenamefont#1{#1}\fi
\expandafter\ifx\csname url\endcsname\relax
  \def\url#1{\texttt{#1}}\fi
\expandafter\ifx\csname urlprefix\endcsname\relax\def\urlprefix{URL }\fi
\providecommand{\bibinfo}[2]{#2}
\providecommand{\eprint}[2][]{\url{#2}}

\bibitem[{\citenamefont{Hammer et~al.}(2020)\citenamefont{Hammer, K\"onig, and
  van Kolck}}]{Hammer:2019poc}
\bibinfo{author}{\bibfnamefont{H.-W.} \bibnamefont{Hammer}},
  \bibinfo{author}{\bibfnamefont{S.}~\bibnamefont{K\"onig}}, \bibnamefont{and}
  \bibinfo{author}{\bibfnamefont{U.}~\bibnamefont{van Kolck}},
  \bibinfo{journal}{Rev. Mod. Phys.} \textbf{\bibinfo{volume}{92}},
  \bibinfo{pages}{025004} (\bibinfo{year}{2020}), \eprint{1906.12122}.

\bibitem[{\citenamefont{Bagnarol et~al.}(2023)\citenamefont{Bagnarol,
  Sch\"afer, Bazak, and Barnea}}]{Bagnarol:2023crb}
\bibinfo{author}{\bibfnamefont{M.}~\bibnamefont{Bagnarol}},
  \bibinfo{author}{\bibfnamefont{M.}~\bibnamefont{Sch\"afer}},
  \bibinfo{author}{\bibfnamefont{B.}~\bibnamefont{Bazak}}, \bibnamefont{and}
  \bibinfo{author}{\bibfnamefont{N.}~\bibnamefont{Barnea}},
  \bibinfo{journal}{Phys. Lett. B} \textbf{\bibinfo{volume}{844}},
  \bibinfo{pages}{138078} (\bibinfo{year}{2023}), \eprint{2306.04036}.

\bibitem[{\citenamefont{Bedaque et~al.}(1999)\citenamefont{Bedaque, Hammer, and
  van Kolck}}]{Bedaque:1998km}
\bibinfo{author}{\bibfnamefont{P.~F.} \bibnamefont{Bedaque}},
  \bibinfo{author}{\bibfnamefont{H.-W.} \bibnamefont{Hammer}},
  \bibnamefont{and} \bibinfo{author}{\bibfnamefont{U.}~\bibnamefont{van
  Kolck}}, \bibinfo{journal}{Nucl. Phys. A} \textbf{\bibinfo{volume}{646}},
  \bibinfo{pages}{444} (\bibinfo{year}{1999}), \eprint{nucl-th/9811046}.

\bibitem[{\citenamefont{Platter et~al.}(2004)\citenamefont{Platter, Hammer, and
  Mei{\ss}ner}}]{Platter:2004he}
\bibinfo{author}{\bibfnamefont{L.}~\bibnamefont{Platter}},
  \bibinfo{author}{\bibfnamefont{H.-W.} \bibnamefont{Hammer}},
  \bibnamefont{and} \bibinfo{author}{\bibfnamefont{U.-G.}
  \bibnamefont{Mei{\ss}ner}}, \bibinfo{journal}{Phys. Rev. A}
  \textbf{\bibinfo{volume}{70}}, \bibinfo{pages}{052101}
  (\bibinfo{year}{2004}), \eprint{cond-mat/0404313}.

\bibitem[{\citenamefont{Bazak et~al.}(2016)\citenamefont{Bazak, Eliyahu, and
  van Kolck}}]{Bazak:2016wxm}
\bibinfo{author}{\bibfnamefont{B.}~\bibnamefont{Bazak}},
  \bibinfo{author}{\bibfnamefont{M.}~\bibnamefont{Eliyahu}}, \bibnamefont{and}
  \bibinfo{author}{\bibfnamefont{U.}~\bibnamefont{van Kolck}},
  \bibinfo{journal}{Phys. Rev. A} \textbf{\bibinfo{volume}{94}},
  \bibinfo{pages}{052502} (\bibinfo{year}{2016}), \eprint{1607.01509}.

\bibitem[{\citenamefont{Ji and Phillips}(2013)}]{Ji:2012nj}
\bibinfo{author}{\bibfnamefont{C.}~\bibnamefont{Ji}} \bibnamefont{and}
  \bibinfo{author}{\bibfnamefont{D.~R.} \bibnamefont{Phillips}},
  \bibinfo{journal}{Few Body Syst.} \textbf{\bibinfo{volume}{54}},
  \bibinfo{pages}{2317} (\bibinfo{year}{2013}), \eprint{1212.1845}.

\bibitem[{\citenamefont{Bazak et~al.}(2019)\citenamefont{Bazak, Kirscher,
  K\"onig, Pav\'on~Valderrama, Barnea, and van Kolck}}]{Bazak:2018qnu}
\bibinfo{author}{\bibfnamefont{B.}~\bibnamefont{Bazak}},
  \bibinfo{author}{\bibfnamefont{J.}~\bibnamefont{Kirscher}},
  \bibinfo{author}{\bibfnamefont{S.}~\bibnamefont{K\"onig}},
  \bibinfo{author}{\bibfnamefont{M.}~\bibnamefont{Pav\'on~Valderrama}},
  \bibinfo{author}{\bibfnamefont{N.}~\bibnamefont{Barnea}}, \bibnamefont{and}
  \bibinfo{author}{\bibfnamefont{U.}~\bibnamefont{van Kolck}},
  \bibinfo{journal}{Phys. Rev. Lett.} \textbf{\bibinfo{volume}{122}},
  \bibinfo{pages}{143001} (\bibinfo{year}{2019}), \eprint{1812.00387}.

\bibitem[{\citenamefont{Carlson et~al.}(2017)\citenamefont{Carlson, Gandolfi,
  van Kolck, and Vitiello}}]{Carlson:2017txq}
\bibinfo{author}{\bibfnamefont{J.}~\bibnamefont{Carlson}},
  \bibinfo{author}{\bibfnamefont{S.}~\bibnamefont{Gandolfi}},
  \bibinfo{author}{\bibfnamefont{U.}~\bibnamefont{van Kolck}},
  \bibnamefont{and} \bibinfo{author}{\bibfnamefont{S.~A.}
  \bibnamefont{Vitiello}}, \bibinfo{journal}{Phys. Rev. Lett.}
  \textbf{\bibinfo{volume}{119}}, \bibinfo{pages}{223002}
  (\bibinfo{year}{2017}), \eprint{1707.08546}.

\bibitem[{\citenamefont{De-Leon and
  Pederiva}(2022)}]{deleon2022equationstatestronglyinteracting}
\bibinfo{author}{\bibfnamefont{H.}~\bibnamefont{De-Leon}} \bibnamefont{and}
  \bibinfo{author}{\bibfnamefont{F.}~\bibnamefont{Pederiva}}
  (\bibinfo{year}{2022}), \eprint{2211.00165}.

\bibitem[{\citenamefont{Stetcu et~al.}(2007)\citenamefont{Stetcu, Barrett, and
  van Kolck}}]{Stetcu:2006ey}
\bibinfo{author}{\bibfnamefont{I.}~\bibnamefont{Stetcu}},
  \bibinfo{author}{\bibfnamefont{B.~R.} \bibnamefont{Barrett}},
  \bibnamefont{and} \bibinfo{author}{\bibfnamefont{U.}~\bibnamefont{van
  Kolck}}, \bibinfo{journal}{Phys. Lett. B} \textbf{\bibinfo{volume}{653}},
  \bibinfo{pages}{358} (\bibinfo{year}{2007}), \eprint{nucl-th/0609023}.

\bibitem[{\citenamefont{Contessi et~al.}(2017)\citenamefont{Contessi, Lovato,
  Pederiva, Roggero, Kirscher, and van Kolck}}]{Contessi:2017rww}
\bibinfo{author}{\bibfnamefont{L.}~\bibnamefont{Contessi}},
  \bibinfo{author}{\bibfnamefont{A.}~\bibnamefont{Lovato}},
  \bibinfo{author}{\bibfnamefont{F.}~\bibnamefont{Pederiva}},
  \bibinfo{author}{\bibfnamefont{A.}~\bibnamefont{Roggero}},
  \bibinfo{author}{\bibfnamefont{J.}~\bibnamefont{Kirscher}}, \bibnamefont{and}
  \bibinfo{author}{\bibfnamefont{U.}~\bibnamefont{van Kolck}},
  \bibinfo{journal}{Phys. Lett. B} \textbf{\bibinfo{volume}{772}},
  \bibinfo{pages}{839} (\bibinfo{year}{2017}), \eprint{1701.06516}.

\bibitem[{\citenamefont{Bansal et~al.}(2018)\citenamefont{Bansal, Binder,
  Ekstr\"om, Hagen, Jansen, and Papenbrock}}]{Bansal:2017pwn}
\bibinfo{author}{\bibfnamefont{A.}~\bibnamefont{Bansal}},
  \bibinfo{author}{\bibfnamefont{S.}~\bibnamefont{Binder}},
  \bibinfo{author}{\bibfnamefont{A.}~\bibnamefont{Ekstr\"om}},
  \bibinfo{author}{\bibfnamefont{G.}~\bibnamefont{Hagen}},
  \bibinfo{author}{\bibfnamefont{G.~R.} \bibnamefont{Jansen}},
  \bibnamefont{and}
  \bibinfo{author}{\bibfnamefont{T.}~\bibnamefont{Papenbrock}},
  \bibinfo{journal}{Phys. Rev. C} \textbf{\bibinfo{volume}{98}},
  \bibinfo{pages}{054301} (\bibinfo{year}{2018}), \eprint{1712.10246}.

\bibitem[{\citenamefont{Yang et~al.}(2021)\citenamefont{Yang, Ekstr\"om,
  Forss\'en, and Hagen}}]{Yang:2020pgi}
\bibinfo{author}{\bibfnamefont{C.-J.} \bibnamefont{Yang}},
  \bibinfo{author}{\bibfnamefont{A.}~\bibnamefont{Ekstr\"om}},
  \bibinfo{author}{\bibfnamefont{C.}~\bibnamefont{Forss\'en}},
  \bibnamefont{and} \bibinfo{author}{\bibfnamefont{G.}~\bibnamefont{Hagen}},
  \bibinfo{journal}{Phys. Rev. C} \textbf{\bibinfo{volume}{103}},
  \bibinfo{pages}{054304} (\bibinfo{year}{2021}), \eprint{2011.11584}.

\bibitem[{\citenamefont{Yang}(2024)}]{Yang:2023wci}
\bibinfo{author}{\bibfnamefont{C.~J.} \bibnamefont{Yang}},
  \bibinfo{journal}{Phys. Rev. C} \textbf{\bibinfo{volume}{109}},
  \bibinfo{pages}{054003} (\bibinfo{year}{2024}), \eprint{2312.05085}.

\bibitem[{\citenamefont{Yang et~al.}(2023)\citenamefont{Yang, Ekstr\"om,
  Forss\'en, Hagen, Rupak, and van Kolck}}]{Yang:2021vxa}
\bibinfo{author}{\bibfnamefont{C.-J.} \bibnamefont{Yang}},
  \bibinfo{author}{\bibfnamefont{A.}~\bibnamefont{Ekstr\"om}},
  \bibinfo{author}{\bibfnamefont{C.}~\bibnamefont{Forss\'en}},
  \bibinfo{author}{\bibfnamefont{G.}~\bibnamefont{Hagen}},
  \bibinfo{author}{\bibfnamefont{G.}~\bibnamefont{Rupak}}, \bibnamefont{and}
  \bibinfo{author}{\bibfnamefont{U.}~\bibnamefont{van Kolck}},
  \bibinfo{journal}{Eur. Phys. J. A} \textbf{\bibinfo{volume}{59}},
  \bibinfo{pages}{233} (\bibinfo{year}{2023}), \eprint{2109.13303}.

\bibitem[{\citenamefont{Contessi
  et~al.}(2024{\natexlab{a}})\citenamefont{Contessi, Pav\'on~Valderrama, and
  van Kolck}}]{Contessi:2024vae}
\bibinfo{author}{\bibfnamefont{L.}~\bibnamefont{Contessi}},
  \bibinfo{author}{\bibfnamefont{M.}~\bibnamefont{Pav\'on~Valderrama}},
  \bibnamefont{and} \bibinfo{author}{\bibfnamefont{U.}~\bibnamefont{van
  Kolck}}, \bibinfo{journal}{Phys. Lett. B} \textbf{\bibinfo{volume}{856}},
  \bibinfo{pages}{138903} (\bibinfo{year}{2024}{\natexlab{a}}),
  \eprint{2403.16596}.

\bibitem[{\citenamefont{Contessi
  et~al.}(2024{\natexlab{b}})\citenamefont{Contessi, Sch\"afer, and van
  Kolck}}]{Contessi:2023yoz}
\bibinfo{author}{\bibfnamefont{L.}~\bibnamefont{Contessi}},
  \bibinfo{author}{\bibfnamefont{M.}~\bibnamefont{Sch\"afer}},
  \bibnamefont{and} \bibinfo{author}{\bibfnamefont{U.}~\bibnamefont{van
  Kolck}}, \bibinfo{journal}{Phys. Rev. A} \textbf{\bibinfo{volume}{109}},
  \bibinfo{pages}{022814} (\bibinfo{year}{2024}{\natexlab{b}}),
  \eprint{2310.15760}.

\bibitem[{\citenamefont{Ekstr\"om and Platter}(2025)}]{Ekstrom:2024dqr}
\bibinfo{author}{\bibfnamefont{A.}~\bibnamefont{Ekstr\"om}} \bibnamefont{and}
  \bibinfo{author}{\bibfnamefont{L.}~\bibnamefont{Platter}},
  \bibinfo{journal}{Phys. Lett. B} \textbf{\bibinfo{volume}{860}},
  \bibinfo{pages}{139207} (\bibinfo{year}{2025}), \eprint{2409.08197}.

\bibitem[{\citenamefont{Suzuki and Varga}({1998})}]{Suzuki:1998bn}
\bibinfo{author}{\bibfnamefont{Y.}~\bibnamefont{Suzuki}} \bibnamefont{and}
  \bibinfo{author}{\bibfnamefont{K.}~\bibnamefont{Varga}},
  \emph{\bibinfo{title}{Stochastic Variational Approach to {Quantum-Mechanical}
  Few-Body Problems}} (\bibinfo{publisher}{Springer Berlin, Heidelberg},
  \bibinfo{year}{{1998}}).

\bibitem[{\citenamefont{K\"onig et~al.}(2016)\citenamefont{K\"onig,
  Grie\ss{}hammer, Hammer, and van Kolck}}]{Konig:2015aka}
\bibinfo{author}{\bibfnamefont{S.}~\bibnamefont{K\"onig}},
  \bibinfo{author}{\bibfnamefont{H.~W.} \bibnamefont{Grie\ss{}hammer}},
  \bibinfo{author}{\bibfnamefont{H.-W.} \bibnamefont{Hammer}},
  \bibnamefont{and} \bibinfo{author}{\bibfnamefont{U.}~\bibnamefont{van
  Kolck}}, \bibinfo{journal}{J. Phys. G} \textbf{\bibinfo{volume}{43}},
  \bibinfo{pages}{055106} (\bibinfo{year}{2016}), \eprint{1508.05085}.

\bibitem[{\citenamefont{Chen et~al.}(1999)\citenamefont{Chen, Rupak, and
  Savage}}]{Chen:1999tn}
\bibinfo{author}{\bibfnamefont{J.-W.} \bibnamefont{Chen}},
  \bibinfo{author}{\bibfnamefont{G.}~\bibnamefont{Rupak}}, \bibnamefont{and}
  \bibinfo{author}{\bibfnamefont{M.~J.} \bibnamefont{Savage}},
  \bibinfo{journal}{Nucl. Phys. A} \textbf{\bibinfo{volume}{653}},
  \bibinfo{pages}{386} (\bibinfo{year}{1999}), \eprint{nucl-th/9902056}.

\bibitem[{\citenamefont{Hackenburg}(2006)}]{Hackenburg:2006qd}
\bibinfo{author}{\bibfnamefont{R.~W.} \bibnamefont{Hackenburg}},
  \bibinfo{journal}{Phys. Rev. C} \textbf{\bibinfo{volume}{73}},
  \bibinfo{pages}{044002} (\bibinfo{year}{2006}).

\bibitem[{\citenamefont{Bedaque et~al.}(2000)\citenamefont{Bedaque, Hammer, and
  van Kolck}}]{Bedaque:1999ve}
\bibinfo{author}{\bibfnamefont{P.~F.} \bibnamefont{Bedaque}},
  \bibinfo{author}{\bibfnamefont{H.-W.} \bibnamefont{Hammer}},
  \bibnamefont{and} \bibinfo{author}{\bibfnamefont{U.}~\bibnamefont{van
  Kolck}}, \bibinfo{journal}{Nucl. Phys. A} \textbf{\bibinfo{volume}{676}},
  \bibinfo{pages}{357} (\bibinfo{year}{2000}), \eprint{nucl-th/9906032}.

\bibitem[{\citenamefont{Wang et~al.}(2012)\citenamefont{Wang, Audi, Wapstra,
  Kondev, MacCormick, Xu, and Pfeiffer}}]{Wang_2012}
\bibinfo{author}{\bibfnamefont{M.}~\bibnamefont{Wang}},
  \bibinfo{author}{\bibfnamefont{G.}~\bibnamefont{Audi}},
  \bibinfo{author}{\bibfnamefont{A.~H.} \bibnamefont{Wapstra}},
  \bibinfo{author}{\bibfnamefont{F.~G.} \bibnamefont{Kondev}},
  \bibinfo{author}{\bibfnamefont{M.}~\bibnamefont{MacCormick}},
  \bibinfo{author}{\bibfnamefont{X.}~\bibnamefont{Xu}}, \bibnamefont{and}
  \bibinfo{author}{\bibfnamefont{B.}~\bibnamefont{Pfeiffer}},
  \bibinfo{journal}{Chin. Phys. C} \textbf{\bibinfo{volume}{36}},
  \bibinfo{pages}{1603} (\bibinfo{year}{2012}).

\bibitem[{\citenamefont{Dilg et~al.}(1971)\citenamefont{Dilg, Koester, and
  Nistler}}]{Dilg:1971dkn}
\bibinfo{author}{\bibfnamefont{W.}~\bibnamefont{Dilg}},
  \bibinfo{author}{\bibfnamefont{L.}~\bibnamefont{Koester}}, \bibnamefont{and}
  \bibinfo{author}{\bibfnamefont{W.}~\bibnamefont{Nistler}},
  \bibinfo{journal}{Phys. Lett. B.} \textbf{\bibinfo{volume}{36}},
  \bibinfo{pages}{208} (\bibinfo{year}{1971}).

\bibitem[{\citenamefont{Adams et~al.}(2021)\citenamefont{Adams, Carleo, Lovato,
  and Rocco}}]{Adams:2020aax}
\bibinfo{author}{\bibfnamefont{C.}~\bibnamefont{Adams}},
  \bibinfo{author}{\bibfnamefont{G.}~\bibnamefont{Carleo}},
  \bibinfo{author}{\bibfnamefont{A.}~\bibnamefont{Lovato}}, \bibnamefont{and}
  \bibinfo{author}{\bibfnamefont{N.}~\bibnamefont{Rocco}},
  \bibinfo{journal}{Phys. Rev. Lett.} \textbf{\bibinfo{volume}{127}},
  \bibinfo{pages}{022502} (\bibinfo{year}{2021}), \eprint{2007.14282}.

\bibitem[{\citenamefont{Gnech et~al.}(2022)\citenamefont{Gnech, Adams, Brawand,
  Carleo, Lovato, and Rocco}}]{Gnech:2021wfn}
\bibinfo{author}{\bibfnamefont{A.}~\bibnamefont{Gnech}},
  \bibinfo{author}{\bibfnamefont{C.}~\bibnamefont{Adams}},
  \bibinfo{author}{\bibfnamefont{N.}~\bibnamefont{Brawand}},
  \bibinfo{author}{\bibfnamefont{G.}~\bibnamefont{Carleo}},
  \bibinfo{author}{\bibfnamefont{A.}~\bibnamefont{Lovato}}, \bibnamefont{and}
  \bibinfo{author}{\bibfnamefont{N.}~\bibnamefont{Rocco}},
  \bibinfo{journal}{Few Body Syst.} \textbf{\bibinfo{volume}{63}},
  \bibinfo{pages}{7} (\bibinfo{year}{2022}), \eprint{2108.06836}.

\bibitem[{\citenamefont{Gnech et~al.}(2024)\citenamefont{Gnech, Fore, Tropiano,
  and Lovato}}]{Gnech:2023prs}
\bibinfo{author}{\bibfnamefont{A.}~\bibnamefont{Gnech}},
  \bibinfo{author}{\bibfnamefont{B.}~\bibnamefont{Fore}},
  \bibinfo{author}{\bibfnamefont{A.~J.} \bibnamefont{Tropiano}},
  \bibnamefont{and} \bibinfo{author}{\bibfnamefont{A.}~\bibnamefont{Lovato}},
  \bibinfo{journal}{Phys. Rev. Lett.} \textbf{\bibinfo{volume}{133}},
  \bibinfo{pages}{142501} (\bibinfo{year}{2024}), \eprint{2308.16266}.

\bibitem[{\citenamefont{Kievsky and Gattobigio}(2016)}]{Kievsky:2015dtk}
\bibinfo{author}{\bibfnamefont{A.}~\bibnamefont{Kievsky}} \bibnamefont{and}
  \bibinfo{author}{\bibfnamefont{M.}~\bibnamefont{Gattobigio}},
  \bibinfo{journal}{Few Body Syst.} \textbf{\bibinfo{volume}{57}},
  \bibinfo{pages}{217} (\bibinfo{year}{2016}), \eprint{1511.09184}.

\bibitem[{\citenamefont{Kievsky et~al.}(2018)\citenamefont{Kievsky, Viviani,
  Logoteta, Bombaci, and Girlanda}}]{Kievsky:2018xsl}
\bibinfo{author}{\bibfnamefont{A.}~\bibnamefont{Kievsky}},
  \bibinfo{author}{\bibfnamefont{M.}~\bibnamefont{Viviani}},
  \bibinfo{author}{\bibfnamefont{D.}~\bibnamefont{Logoteta}},
  \bibinfo{author}{\bibfnamefont{I.}~\bibnamefont{Bombaci}}, \bibnamefont{and}
  \bibinfo{author}{\bibfnamefont{L.}~\bibnamefont{Girlanda}},
  \bibinfo{journal}{Phys. Rev. Lett.} \textbf{\bibinfo{volume}{121}},
  \bibinfo{pages}{072701} (\bibinfo{year}{2018}), \eprint{1806.02636}.

\bibitem[{\citenamefont{Gattobigio et~al.}(2019)\citenamefont{Gattobigio,
  Kievsky, and Viviani}}]{Gattobigio:2019omi}
\bibinfo{author}{\bibfnamefont{M.}~\bibnamefont{Gattobigio}},
  \bibinfo{author}{\bibfnamefont{A.}~\bibnamefont{Kievsky}}, \bibnamefont{and}
  \bibinfo{author}{\bibfnamefont{M.}~\bibnamefont{Viviani}},
  \bibinfo{journal}{Phys. Rev. C} \textbf{\bibinfo{volume}{100}},
  \bibinfo{pages}{034004} (\bibinfo{year}{2019}), \eprint{1903.08900}.

\bibitem[{\citenamefont{Deltuva et~al.}(2020)\citenamefont{Deltuva, Gattobigio,
  Kievsky, and Viviani}}]{Deltuva:2020aws}
\bibinfo{author}{\bibfnamefont{A.}~\bibnamefont{Deltuva}},
  \bibinfo{author}{\bibfnamefont{M.}~\bibnamefont{Gattobigio}},
  \bibinfo{author}{\bibfnamefont{A.}~\bibnamefont{Kievsky}}, \bibnamefont{and}
  \bibinfo{author}{\bibfnamefont{M.}~\bibnamefont{Viviani}},
  \bibinfo{journal}{Phys. Rev. C} \textbf{\bibinfo{volume}{102}},
  \bibinfo{pages}{064001} (\bibinfo{year}{2020}), \eprint{2011.06828}.

\bibitem[{\citenamefont{Gattobigio and Kievsky}(2023)}]{Gattobigio:2023fmo}
\bibinfo{author}{\bibfnamefont{M.}~\bibnamefont{Gattobigio}} \bibnamefont{and}
  \bibinfo{author}{\bibfnamefont{A.}~\bibnamefont{Kievsky}},
  \bibinfo{journal}{Few Body Syst.} \textbf{\bibinfo{volume}{64}},
  \bibinfo{pages}{86} (\bibinfo{year}{2023}), \eprint{2305.16814}.

\bibitem[{\citenamefont{Kirscher et~al.}(2010)\citenamefont{Kirscher,
  Griesshammer, Shukla, and Hofmann}}]{Kirscher:2010dgl}
\bibinfo{author}{\bibfnamefont{J.}~\bibnamefont{Kirscher}},
  \bibinfo{author}{\bibfnamefont{H.~W.} \bibnamefont{Griesshammer}},
  \bibinfo{author}{\bibfnamefont{D.}~\bibnamefont{Shukla}}, \bibnamefont{and}
  \bibinfo{author}{\bibfnamefont{H.~M.} \bibnamefont{Hofmann}},
  \bibinfo{journal}{Eur. Phys. J. A} \textbf{\bibinfo{volume}{44}},
  \bibinfo{pages}{239} (\bibinfo{year}{2010}), \eprint{0903.5538}.

\bibitem[{\citenamefont{Schiavilla et~al.}(2021)\citenamefont{Schiavilla,
  Girlanda, Gnech, Kievsky, Lovato, Marcucci, Piarulli, and
  Viviani}}]{Schiavilla:2021dun}
\bibinfo{author}{\bibfnamefont{R.}~\bibnamefont{Schiavilla}},
  \bibinfo{author}{\bibfnamefont{L.}~\bibnamefont{Girlanda}},
  \bibinfo{author}{\bibfnamefont{A.}~\bibnamefont{Gnech}},
  \bibinfo{author}{\bibfnamefont{A.}~\bibnamefont{Kievsky}},
  \bibinfo{author}{\bibfnamefont{A.}~\bibnamefont{Lovato}},
  \bibinfo{author}{\bibfnamefont{L.~E.} \bibnamefont{Marcucci}},
  \bibinfo{author}{\bibfnamefont{M.}~\bibnamefont{Piarulli}}, \bibnamefont{and}
  \bibinfo{author}{\bibfnamefont{M.}~\bibnamefont{Viviani}},
  \bibinfo{journal}{Phys. Rev. C} \textbf{\bibinfo{volume}{103}},
  \bibinfo{pages}{054003} (\bibinfo{year}{2021}), \eprint{2102.02327}.

\bibitem[{\citenamefont{Bub et~al.}(2024)\citenamefont{Bub, Piarulli,
  Furnstahl, Pastore, and Phillips}}]{Bub:2024gyz}
\bibinfo{author}{\bibfnamefont{J.~M.} \bibnamefont{Bub}},
  \bibinfo{author}{\bibfnamefont{M.}~\bibnamefont{Piarulli}},
  \bibinfo{author}{\bibfnamefont{R.~J.} \bibnamefont{Furnstahl}},
  \bibinfo{author}{\bibfnamefont{S.}~\bibnamefont{Pastore}}, \bibnamefont{and}
  \bibinfo{author}{\bibfnamefont{D.~R.} \bibnamefont{Phillips}}
  (\bibinfo{year}{2024}), \eprint{2408.02480}.

\end{thebibliography}

\end{document}